\documentclass[traditabstract,letter]{aa}
\usepackage[varg]{txfonts}
\usepackage{graphicx}
\usepackage{subfigure}
\usepackage{rotating}
\usepackage{xcolor}
\usepackage{ragged2e}
\usepackage{soul} 
\usepackage{float}
\usepackage{amsmath}
\usepackage{wrapfig,lipsum}
\usepackage[flushleft]{threeparttable} 
\usepackage{multirow}
\DeclareUnicodeCharacter{2212}{-}
\usepackage{arydshln}
\usepackage{tikz}
\usepackage{comment}
\excludecomment{toexclude} 
\definecolor{applegreen}{rgb}{0.55, 0.71, 0.0}

\definecolor{pythonpurple}{RGB}{128, 0, 128}
\definecolor{pythongreen}{RGB}{0, 128, 0}
\definecolor{pythonyellowgreen}{RGB}{154, 205, 50}
\definecolor{pythonroyalblue}{RGB}{65, 105, 225}
\definecolor{pythonforestgreen}{RGB}{34, 139, 34}
\definecolor{pythondarkred}{RGB}{139, 0, 0}

\bibpunct{(}{)}{;}{a}{}{,} 

\begin{document}
\title{M31's nucleus: molecular and ionised gas content upper limits}
\author{
          Anne-Laure Melchior\inst{1}
          \and
          Fran\c coise Combes\inst{1}\fnmsep\inst{2}
          }

\institute{LERMA, Sorbonne Universit\'{e}, Observatoire de Paris, Universit\'{e} PSL, CNRS, F-75014, Paris, France\\
              \email{Anne-Laure.Melchior@observatoiredeparis.psl.eu}
         \and
             Coll\`{e}ge de France, 11 Place Marcelin Berthelot, 75005 Paris, France\\
}



\abstract{We report NOEMA and ALMA observations of the nucleus of Andromeda (M31), putting strong constraints on the presence of gas in the form of cold or warm phase, as proposed by Chang et al. M31 hosts the largest supermassive black hole (SMBH) closer than 1\,Mpc from us. Its nucleus is  silent with some murmurs at the level of $ 4 \times 10^{-9} L_{Edd}$, and is surrounded by a 5-pc-radius disk of old stars. The mass-loss from these stars is expected to fill a molecular gas disk within the tidal truncation of 1 pc ($=0.26$\,arcsec), of 10$^4 M_\odot$, corresponding to a CO(1-0) signal of 2\,mJy with a linewidth of 1000 km/s. 
We observed the nucleus with NOEMA in CO(2-1) and with ALMA in CO(3-2) with angular resolutions of $0.5$\arcsec (1.9~pc) and $0.12$\arcsec (0.46~pc) respectively. We exclude the presence of gas with a 3$\sigma$ upper limit of 195$M_\odot$.
The CO(3-2) upper limit also constrains warm gas, escaping detection in CO(1-0). The scenario proposed by Chang et al. is not verified, and instead the hot gas, expelled by the stellar winds, might never cool nor fall onto the disc. {Alternatively, the stellar wind mass-loss rate can have been overestimated by a factor $50$, and/or the ionised gas escaped from the nucleus.}
\\
The SMBH in M31 is obviously in a low state of activity, similar to what is observed for Sgr A* in the Milky Way (MW). Recently, a cool (10$^4$ K) ionised accretion disc has been detected around Sgr A* in the H30$\alpha$ recombination line with ALMA. Re-scaling sizes, masses and fluxes according to the mass of M31's black hole (35 times higher than in the MW) and the distances, a similar disc could be easily detectable around M31 nucleus with an expected signal 8 times weaker that the signal detected in SgrA*.  We searched for an ionised gas disc around M31 nucleus with NOEMA, and put a 3$\sigma$ upper limit on the H30$\alpha$ recombination line at a level twice lower than  expected with a simple scaling of the SgrA*.} 

\maketitle

\section{Introduction}
M31 hosts the closest SMBH, more massive than SgrA*, with a mass of $1.4 \times 10^8\,M_\odot$ \citep{2005ApJ...631..280B}. 
While this black hole has probably been active in the past \citep{2016MNRAS.459L..76P,2019RAA....19...46F,Zhang-Foster2019}, it actually murmurs at the level of $4\times 10^{−9}\,L_{Edd}$ \citep{2011ApJ...728L..10L}. The M31 galaxy is known to be in the green valley \citep{1996ASPC..103..241H,2014ARA&A..52..589H}, which is supported by the very little amount of molecular gas ($\sim 8 \times 10^4 M_\odot$) detected inside 250\,pc 
\citep{2000MNRAS.312L..29M,2011A&A...536A..52M,2013A&A...549A..27M,2016A&A...585A..44M,2017A&A...607L...7M,2019A&A...625A.148D}, and by the very little star formation ($\sim 10^{−5}\,M_\odot$ yr$^{−1}$) observed in the central kpc \citep{2015MNRAS.451.4126D,2018MNRAS.478.5379D,2018AJ....156..269L,2022AJ....163..138L}. The  molecular gas detected in the central kpc is very clumpy and its kinematics irregular suggesting that the detected gas might be seen in projection \citep{2019A&A...625A.148D}. These arguments are compatible with an exhaustion of gas inside-out \citep{2016MNRAS.455.1218B}.
The high resolution photometry performed within the sphere of influence of the M31 black hole revealed several peaks in intensity: as discussed in \citet{2005ApJ...631..280B}, the nucleus reveals three components. This has been explained by an eccentric disc \citep{1995AJ....110..628T}, extending at $\sim$ 4~pc from the black hole. It is composed of old stars with a bright stellar concentration in P1 at apoapse, and a fainter concentration in P2, at periapse. In this scheme, the third peak designated as P3 is a nuclear blue stellar cluster located next to the black hole \citep{2012ApJ...745..121L}. Contrarily to the MW hosting Wolf-Rayet stars close to SgrA* \citep[e.g.][]{2022ApJ...932L..21M}, P3 has been interpreted as a compact cluster of young stars thought to be 100-200\,Myr old. Such central young stellar population is also detected in nearby galaxies \citep{2006AJ....132.1074R,2006AJ....132.2539S} and in the Galactic centre \citep{2006ApJ...643.1011P}. While the formation of nuclear stellar clusters in massive galaxies is usually associated to central star formation \citep{2020A&ARv..28....4N}, it has also been discussed that these centrally concentrated blue stars could be blue straggler stars, extended horizontal branch stars and/or young, recently formed, stars \citep{2016MNRAS.463.1605L}. Indeed, different mechanisms could drive gas infall onto the centre to form new stars, or accrete onto old main sequence stars already present. The non-detection of molecular gas next to the black hole is closely related to the coevolution of these nuclear clusters with the central black hole \citep[e.g.][]{2020A&ARv..28....4N}.

\citet{2010MNRAS.407.1529H,2010MNRAS.405L..41H} have shown that eccentric nuclear stellar discs may originate in gas-rich galaxy mergers. Relying on numerical simulations, they show that M31's nuclear stellar disc can well be reproduced by gas-rich accretion onto the black hole \citep{2010MNRAS.405L..41H}. This mechanism could have accounted for the growth of the SMBH, {for} which the existing disc is a stable relic of this past active phase.   \citet{2018ApJ...853..141M} relied on N-body simulations to show that the smearing of the stellar orbits due to differential precession does not occur, as a torque of the orbit adds to the precession and stabilizes the eccentric disc. 
In M31, \citet{2007ApJ...668..236C} have proposed that the nuclear eccentric old-stellar disc \citep{1995AJ....110..628T} could be cyclically replenished from the gas expelled through mass loss of red giants and asymptotic giant branch stars.  Gas on orbits crossing the tidal radius $R_t$ would collide, shock and fall into a closer orbit around the black hole. The gas would accumulate into a nuclear disc until star formation events are triggered every 500\,Myr. According to \citet{2007ApJ...668..236C}, the key parameter and source of uncertainty of this modelling is the precession rate of the eccentric disc $\Omega_P$, which should not exceed 3-10\,km\,s$^{−1}$/pc, which is in agreement with the overall argument that the central eccentric disc is long-lived \citep{2018ApJ...853..141M}. \citet{Bacon2001} found from simulations a natural m=1 mode in the nuclear disc, with a very slow pattern speed (3\,km\,s$^{−1}$/pc) that can be maintained during more than a thousand dynamical times. \citet{2018ApJ...854..121L} has estimated with OSIRIS/Keck spectroscopy $\Omega_P = 0.0 \pm 3.9$\,km\,s\,$^{−1}$/pc, which is compatible with the results of \citet{2010MNRAS.405L..41H}.

In \citet{2013A&A...549A..27M}, we first observed the nucleus at IRAM-30m with 12\,arcsec resolution in CO(2-1). No molecular gas associated with the black hole has been detected. Only some gas mass further out, within 100\,pc, corresponding to about $4.2 \times 10^4 M_\odot$ has been detected. With these single-dish observations, an rms sensitivity of 20\,mJy for a velocity resolution of 2.6\,km\,s$^{-1}$ has been reached. The nucleus has been subsequently observed with NOEMA in CO(1-0) \citep{2017A&A...607L...7M}. 
A small clump of $2000\,M_\odot$ with
$\Delta v = 14$ km\,s$^{−1}$ within 9\,pc from the centre, most probably seen in projection, has been detected. At the black hole position, an rms sensitivity of 3.2\,Jy/beam
with a beam of 3.37$^{\prime\prime} \times$2.44$^{\prime\prime}$ has been reached for a velocity resolution of 5.1\,km\,s$^{−1}$, which corresponds to a $3\sigma$ upper limit on the molecular gas mass of $4300\,M_\odot$ for a linewidth of 1000\,km\,s$^{−1}$. This rules out the original prediction of \citet{2007ApJ...668..236C}, namely a CO(1-0) flux of 2\,mJy with a linewidth of 1000\,km\,s$^{−1}$ corresponding to a molecular mass of $10^4\,M_\odot$ gas concentrated inside the tidal truncation radius $R_t < 1$\,pc.
However, we do expect stellar mass loss and winds from the old stellar population of the eccentric disc, and hence an accumulation of gas in the centre. It is hence possible that previous NOEMA observations with a $3.37^{\prime\prime} \times 2.44^{\prime\prime}$ (i.e. about 13\,pc $\times$ 9\,pc) beam did not succeed to detect this gas by lack of sensitivity, due to the dilution of the signal. In this paper, we present new NOEMA and ALMA observations of M31’s centre, gaining an order of magnitude in sensitivity and resolution.  

As M31's black hole is not in an active phase (like SgrA*), given the lack of gas in the nuclear region and its low Bondi accretion rate \citep[\protect$\dot{M_{B}} \sim 7 \times 10^{-5} M_{\odot}\,yr^{-1}$;][]{2010ApJ...710..755G}, it is probably typical of the so-called radiatively inefficient accretion flows due to very low density hot gas \citep[$n_e = 0.10\pm0.04$\,cm$^{-3}$][]{2002ASPC..262..147D,Inayoshi2018}.  In such a configuration with low Eddington luminosity \citep[$10^{-9} L_\odot$;][]{2010ApJ...710..755G},  the gas does not cool via radiation, and the hot accretion disc is probably thick with advection-dominated accretion flows (ADAF) \citep[e.g.][]{1997ApJ...482..400E,2004A&A...413..861M}.  In X-ray, the detected gas has a typical temperature of 0.3\,keV $\sim 6\times 10^6$\,K \citep{2002ASPC..262..147D}. The cooling process in this multi-phase region is complex and is expected to proceed with fragmentation
\citep{2018MNRAS.473.5407M}.

Such low states are observed in nearby nuclei, as in the Galactic Centre,
where M$_\bullet$= 4 10$^6$ M$_\odot$, and L$_{bol}$= 2 10$^{-9}$ L$_{Edd}$, or M31,
where  M$_\bullet$= 1.2 10$^8$ M$_\odot$, and L$_{bol}$= 10$^{-9}$ L$_{Edd}$
\citep[e.g.][]{Inayoshi2018}. Near the nucleus, there should exist
a rotating disc, expected to be geometrically thick and optically thin,
where the convection takes the energy away and limits the accretion.
This can explain the low luminosity, due to the
radiatively inefficient accretion flow (RIAF).

Recently, \citet{2019Natur.570...83M} have detected the recombination line H30$\alpha$
in the accretion disc around the Galactic Centre with ALMA (see their figure 1).
Although their beam is $\sim$ 0.3\,arcsec, they are able to see a rotating disc, with the
blue and red-shifted sides peaking each 0.11\,arcsec from the centre (i.e. 0.004\,pc); this is 1/10th of the Bondi radius R$_B$, or 10$^4$ of the horizon radius. The width of the line (2200\,km\,s$^{-1}$) corresponds to the rotation around a black hole mass of $4\times 10^6$ M$_\odot$, at a radius R$_B$/10. The gas mass corresponds to $\sim$ 10$^{-4.5}$ M$_\odot$, with an average density of 4000 cm$^{-3}$, with may be
some clumps at n=10$^5$-10$^6$ cm$^{-3}$. The emission measured is proportional to n$^2$ times the volume occupied by the ionised gas. It has been amplified by the millimetre continuum of SgrA* by a factor 80.
This masing effect is likely to occur also in M31.

{Throughout this paper, we consider a luminosity distance of 0.78 Mpc  for M31 \citep[e.g.][]{1998ApJ...503L.131S}, corresponding to $1^{\prime\prime}=3.8$\,pc.} In Sect. \ref{sect:obs}, we describe the observations we carried out to search for CO and H30$\alpha$ recombination lines next to the SMBH. In Sect. \ref{sect:upper}, we present the upper limits thus achieved. In Sect. \ref{sect:disc}, we discuss our results.

\section{Observations and upper limits}
\label{sect:obs}
We describe below the set of each (sub)millimeter observations carried out with the phase center position at the optical position \citep{1992ApJ...390L...9C}, namely $00^h42^m44.37^s\,+41^d16^m8.34^s$. The upper limits provided here have been derived for this position. Nevertheless, we check that given the size of the primary beams (22\arcsec and 18\arcsec respectively) this does not change the result at the position of the radio source $00^h42^m44.33^s\,+41^d16^m08.42^s$, as both are separated by 0.6\arcsec.

\subsection{NOEMA observations}
The first epoch observations (2012) of CO(1-0) have been described in \citet{2017A&A...607L...7M} and \citet{2019A&A...625A.148D}.

{In 2020, we observed the nucleus at 231\,GHz with the A configuration, reaching a 0.3\,arcsec, with 10 antennas and 2 polarisations. We thus targeted the H30$\alpha$ recombination line. We reached an rms noise level of 0.11 (resp. 0.074)\,mJy in 8 hours of integration time in 1000 (resp. 2200)\,km\,s$^{-1}$ channels. }

We also get limits on the CO(2-1) line at 230.769\,GHz. The primary beam diameter was 22\,arcsec (88\,pc). We applied standard calibrations with pointing and tuning on IRAM calibrators (3C454.3, MWC349, 0010+405, 0003+380). 
\begin{figure}[h!]
    \centering
    \includegraphics[width=\linewidth]{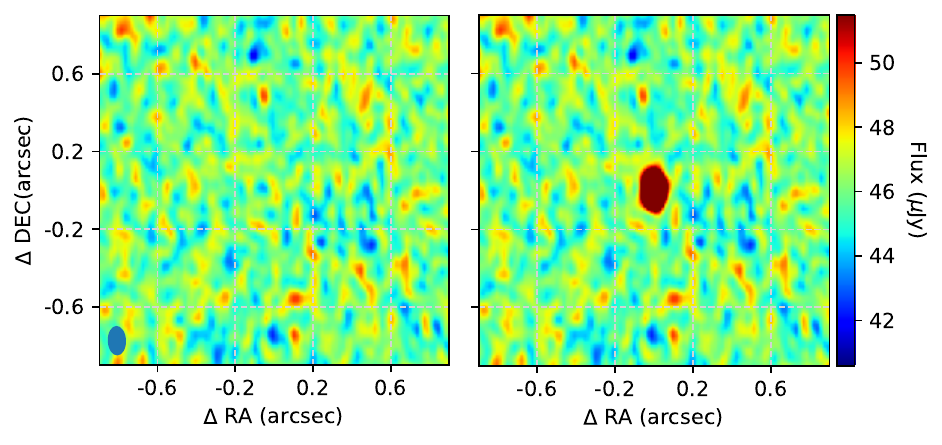}
    \caption{Noise level obtained with ALMA observation of M31's nucleus in the CO(3-2) frequency range. The left panel displays the noise level achieved for $\Delta v=1000$\,km/s. The beam is presented in the lower left corner. In the right panel, the $2$\,mJy CO(1-0) signal expected by \citet{2007ApJ...668..236C}, assuming $R_{13}=1.8$ has been superimposed on the noise map.}
    \label{fig:ALMAmap}
\end{figure}
\begin{table}[h]
\caption{Summary of the different line observations {{($3\sigma$ upper limits)}}. We provide for each observational set-up the characteristics of the beam and the integration time $t_{int}$, for the instruments NOEMA (N) and ALMA (A). The integrated line ${S\Delta V|}^{3\sigma}_{up}$ has been computed assuming a $\Delta V_{max}=$\,1000\,km/s window centred on the systemic velocity for the lines with the 99.7$\%$  CL upper limit estimates.
}
\centering
\begin{tabular}{l|cc|ccc}
\hline
Inst. & \multicolumn{1}{c}{Beam}     & t$_{int}$&  $\nu_{obs}$   & Line          &  \multicolumn{1}{c}{${S\Delta V|}^{3\sigma}_{up}$}  \\\hline
           & \multicolumn{1}{c}{$" \times "$,$\deg$}& & GHz           &    CO          &Jy\,km/s \\\hline
N$^a$     &2.93$,$2.05$,$-2.9    & 6$^+$& 115.387 & (1-0)  & 1.32\\ 
N$^b$     &0.60$,$0.42$,$-164.   & 8.13 & 230.769 & (2-1)  &  0.33 \\
N$^b$     &0.66$,$0.46$,$+136.   & 8.13 & 232.133 & H30$\alpha$ & 0.99 \\
A$^c$      &0.15$,$0.09$,$2.3    & 1.32 & 346.142 &  (3-2) &0.15 \\
 \hline
\end{tabular}
\tablefoot{
The upper limits on HCO+(4-3) at 356.734\,GHz are the same as those estimated for CO(3-2).\\
$^+$ This was a mosaic of 4 fields, each integrated 6 hours \citep[][]{2017A&A...607L...7M}.\\
Observation epochs: $^a$: between 3 September and 24 December 2012;  $^b$: 8$^{th}$-14$^{th}$ February 2020; $^c$: July 2021}
\label{tab:Line_upperlimits}
\end{table}

\subsection{ALMA observations}
{We observed the nucleus at 346.142\,GHz (band 7) targetting the CO(3-2) line with the ALMA-12m array in two configurations, within the project 2019.1.00711.S (PI: Melchior). We reached an angular resolution of 0.12\arcsec (resp. 0.8\arcsec) corresponding to the baseline configuration C43.6 (resp. C43.3), with 1.3 (resp. 0.33) hours of integration time. We also have the HCO$^+$(4-3) line, in the upper sideband. The primary beam diameter was 18\,arcsec (88\,pc). The standard pipeline has been applied.} { Figure \ref{fig:ALMAmap} displayed a 1.6$^{\prime\prime}\times$1.6$^{\prime\prime}$ (6\,pc$\times$6\,pc) field of view centred on M31's black hole. While the left panel displays the noise level achieved for a $\Delta V = 1000$\,km/s bandwidth, the right panel shows the signal expected according to \citet{2007ApJ...668..236C}, assuming $R_{13}=1.8$ (as defined in Sect. \ref{ssect:H2}).}

\section{Upper limits}
\label{sect:upper}
\begin{figure}
	\centering
\includegraphics[width=.49\textwidth]{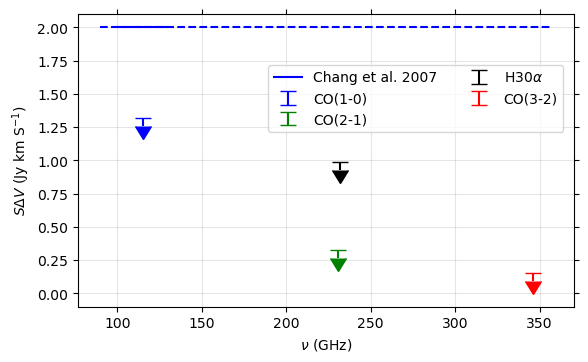}
\caption{Upper limits ($3\,\sigma$) on the line fluxes. The tick positions correspond to the 3\,$\sigma$ upper values. The CO (resp. H30$\alpha$) line measurements are displayed in blue, green and red (resp. black). The prediction of \citet{2007ApJ...668..236C} is displayed with the horizontal blue line.}
\label{fig:lineupper}
\end{figure}

\begin{figure}[h]
	\centering
\includegraphics[width=.49\textwidth]{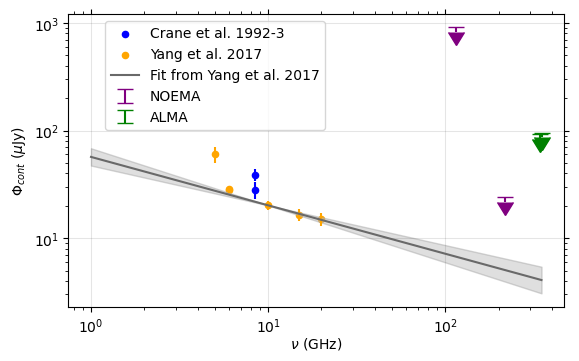}
\caption{Upper limits on the continuum flux. The blue and orange points correspond to the measurements of \citet{1992ApJ...390L...9C,1993ApJ...417L..61C} and \citet{2017ApJ...845..140Y}. The arrows correspond to the upper limits. The ticks correspond to the 3\,$\sigma$  upper limit values. The line and associated error bars correspond to the fit discussed in \citet{2017ApJ...845..140Y}.}
\label{fig:contupper}
\end{figure}

Table \ref{tab:Line_upperlimits} summarises the $3\sigma$ upper limits achieved on the observed line intensities, together {with} the beam, integration time $t_{int}$ and their observed frequencies, while Table \ref{tab:Cont_upperlimits} gives the $3\sigma$ upper limits on the continuum fluxes. {{
Similarly, Figures \ref{fig:lineupper} and \ref{fig:contupper} illustrate the upper limits reached with the measurements described in the previous section. 
In the following, we discuss the significance of these three types of upper limits, namely on the molecular gas lines, the radio recombination line and the continuum.}}
\begin{table}[h!]
\caption{{{Summary of the different continuum observations ($3\sigma$ upper limits)}}. The 99.7$\%$ CL upper limit on the continuum flux (${\sigma^{{{max}}}_{up}}$) computed for $\Delta V =1000$\,km\,s$^{-1}$ at the central band frequency $\nu_{obs}$ on the whole available band $\Delta \nu$. The low frequency side of the ALMA band was more sensitive.
}
\centering
\begin{tabular}{l|lll}
\hline
Instrument & \multicolumn{1}{c}{$\nu_{obs}$}   &  \multicolumn{1}{c}{$\Delta \nu$}& \multicolumn{1}{c}{${3\sigma^{ {{max}}}_{up}}$} \\\hline
           &  \multicolumn{1}{c}{GHz}  &   GHz & $\mu$Jy/beam\\\hline
NOEMA & 114.739 & 3.6 & 912.\\
NOEMA & 218.327 & 7.5 & 24.\\
ALMA & 351.071 & 7.5 & 96.$^{++}$\\
ALMA  & 344.332; 351.463 & 1.8; 7.3 & 93.; 150.\\
 \hline
\end{tabular}
\tablefoot{
$^{++}$ This estimate has been derived on the spatial standard deviation of the stacked continuum image produced in the standard data reduction, computed on a area covering 10 times the beam.
}
\label{tab:Cont_upperlimits}
\end{table}

\subsection{Molecular gas}
\label{ssect:H2}
To derive upper limits on the total H$_2$ mass, we compute the intrinsic CO luminosity with the velocity integrated transition line flux ${\rm F_{CO(J \rightarrow J-1)}} (\propto {\rm S \Delta V})$ and calculate\,:
\begin{equation}\label{eq:l_co}
     {\rm \left( \frac{L'_{CO(J \rightarrow J-1)}}{K \, km \, s^{-1} \, pc^{2}} \right) = 3.25 \times 10^{7} \left( \frac{F_{CO(J \rightarrow J-1)}}{Jy \, km \, s^{-1}} \right)  \left( \frac{\nu_{rest}}{GHz} \right)^{-2}  \left( \frac{D_{L}}{Mpc} \right)^{2}},
\end{equation}
where $\nu_{\rm rest}$ is the rest CO line frequency and ${\rm D_L}$ the luminosity distance \citep{1997ApJ...478..144S}. We consider a luminosity distance of 0.78 Mpc  for M31 \citep[e.g.][]{1998ApJ...503L.131S}. We  then derive the total molecular-gas mass including a correction of $36\,\%$ for interstellar helium using\,:
\begin{equation}\label{eq:co2mh2}
    {\rm M_{H_2} = \alpha_{CO} R_{1J} \, L'_{CO(J \rightarrow J-1)} },
\end{equation}
where the mass-to-light ratio $\alpha_{\rm CO}$ denotes the CO(1-0) luminosity-to-molecular-gas-mass conversion factor, { and  ${\rm R_{1J} = L^{\prime}_{CO\,(1\rightarrow 0)} / L^{\prime}_{CO\,(J \rightarrow J-1)}}$ is} the CO line ratio. 
{\citet{2018ApJ...853..179T} assume {${\rm R_{12}\sim 1.16-1.3}$}  { and ${\rm R_{13}\sim 1.8}$} for star-formation  main-sequence galaxies. One can note that the so-called CO-ladder ratios strongly depend on the column density and the gas temperature \citep{1999A&A...345..369C,2007ASPC..375...25W}. Given the uncertainties here, we consider in the following standard values ${\rm R_{12}\sim 1.3}$ and ${\rm R_{13}\sim 1.8}$.}
For $\alpha_{\rm CO}$, given the absence of gas, we assume a Galactic value $\alpha_{\rm CO}=4.36{\rm M_\odot/(K\,km\,s^{-1}\,pc^2)}$, {{which includes the usual correction to account for helium} \citep[e.g.][]{1987ApJ...319..730S,2010MNRAS.407.2091G,2013ARA&A..51..207B}. {Again, we might consider that the gas in the central part is expected to have a high metallicity. However, the various works \citep[e.g.][]{2011ApJ...741...12B,2012ApJ...746...69G} arguing for a lower  $\alpha_{\rm CO}$ for high-metallicity gas are based on actively star-forming galaxies. Alternatively, it is difficult to argue for a higher  $\alpha_{\rm CO}$, as expected in low-metallicity regions like in the outskirts of galaxies. Hence, upper values derived on the molecular gas are provided for the sake of discussion to compare with standard MW values. As further discussed in Sect. \ref{sect:disc}, it is possible that the molecular gas in this region is CO-dark, due to special physical conditions.}}
The $3\sigma$ upper limits on the molecular mass derived from our measurements are provided in Table \ref{tab:h2mass}. 
The strongest constraint can be derived from the ALMA observations in CO(3-2). 
We can thus exclude that there are more than $195\,M_\odot$ of molecular hydrogen within 1\,pc of the nucleus, {assuming standard MW-like conditions}. {This value is a factor $50$ lower that the mass of gas expected in the \citet{2007ApJ...668..236C} modelling}.
\begin{table}
\caption{Upper limits (3$\sigma$) on the intrinsic CO luminosity and on the molecular hydrogen mass.}
\begin{tabular}{lll|rr}
\hline
Line &$\nu_{rest}$ &  ${\rm F_{CO}}$ & ${\rm L^\prime_{CO}}$ & $M_{H_2}$\\
     & GHz        &  Jy\,km\,s$^{-1}$ & K\,km\,s$^{-1}$\,pc$^2$ & $M_{\odot}$\\\hline
CO(1-0) & 115.271 & 1.32 & 1964 & 8565\\
CO(2-1) & 230.538 & 0.33 & 123 & 696\\
CO(3-2) & 345.796 & 0.15 & 24.5 & 195\\\hline
\end{tabular}
\label{tab:h2mass}
\end{table}
\begin{table*}[t]
\caption{Comparison {between the observed MW's \citep{2019Natur.570...83M} and the expected M31's ionised gas discs, based on their relative black hole masses M$_\bullet$ and distances $D$.} R$_S$ and R$_B$ are the Schwarschild and Bondi radii (in columns 3 to 5), {scaling with the black hole masses M$_\bullet$ (column 2)}. The same value of width of the line $\Delta$V (in column 6)  as observed for the MW is expected for M31. The mass of the {ionised} gas disc {(M$_{gas}$, in column 7) is expected proportional to R$_B^2$, scales with $M^2_\bullet$}. {Therefore, the $H30\alpha$ flux in M31 is expected {8 times smaller than toward} the MW.  Column 8  provides the fluxes expected by simple scaling ratio with respect to the MW detections, while column 9 gives the actual measurement for the MW \citep{2019Natur.570...83M} and the effective upper limit achieved in this paper.}
{The last line summarises the ratio of all the quantities expected (or known) for M31 with respect to the MW. Quantities in italic correspond to predicted values.}
}
\centering
\begin{tabular}{l|ccccccc|cc}
\hline
Galaxy & D & M$_\bullet$ & R$_S$ & R$_B$ & R$_B$ & $\Delta$V&M$_{gas}$& S$_{H30\alpha}^{pred}$&
S$_{H30\alpha}^{obs}$  \\
    & kpc & 10$^6$ M$_\odot$ & pc & pc& '' & km/s & M$_\odot$ & mJy& mJy \\
\hline
& (1) & (2) & (3) & (4) & (5) & (6) & (7) & (8) & (9)   \\
\hline
MW & 8   & 4   & 4 10$^{-7}$   & {0.12}& 1   & 2200 & 10$^{-4.5}$& {\em 3} &3     \\
M31& 780 & 140 & 1.4 10$^{-5}$ &{4.2}  & 0.3 & {\em 2200} & {\em 0.04}        &{\em 0.38} &$< 0.22$ ($3\,\sigma$)  \\\hline
M31/MW Ratio & 97 & 35  & 35   & 35  & 0.3 & {\em 1}    & {\em 1225} &{{\em 0.13}} &0.07   \\
\hline
\end{tabular}
\label{tab:rrl}
\end{table*}

\subsection{Hydrogen recombination line}
In Table \ref{tab:rrl}, {we summarise our results given our initial assumptions scaled from the previous observations of the H30$\alpha$ recombination line detected in the MW for SgrA* by \citet{2019Natur.570...83M}. The Bondi radius scales with the black hole mass $R_B = 2 G M_\bullet/c^2_s$ (where $c_s$ is the
sound speed of the gas): the Bondi radius is thus a priori 35 times larger around M31's black hole than for SgrA* for a given sound speed\,:
\begin{equation}
     R_{B}|_{M31} =  \dfrac{M_{\bullet}|_{M31}}{M_{\bullet}|_{MW}} \times R_{B}|_{MW}  = 35 \times R_{B}|_{MW} 
\end{equation}
Relying on \citet{2003ApJ...591..891B}, who estimated $c_s = 550\,$km\,s$^{-1}$, we found a typical size of 4.2\,pc, which is well sampled by our observations (0.3" or 1.1\,pc resolution).

In addition, the line width $\Delta V$ traces the Keplerian velocity around the black hole and scales as  $\sqrt{M_\bullet/R_B}$. Hence, given the previous assumption on the sound speed, we expect the same line width $V_{H30\alpha}|_{M31}$ for M31 than for the MW\,:
\begin{eqnarray}
     \Delta V_{H30\alpha}|_{M31} &=&  \sqrt{\dfrac{M_{\bullet}|_{M31}}{M_{\bullet}|_{MW}}}\sqrt{\dfrac{R_{B}|_{MW}}{R_{B}|_{M31}}}\times \Delta V_{H30\alpha}|_{MW} \notag \\
     \Delta V_{H30\alpha}|_{M31} &=& \Delta V_{H30\alpha}|_{MW} \sim  2200\,\mathrm{km}\mathrm{s}^{-1}
\end{eqnarray}
}
{Indeed,} given the similarities between the two nuclei,  {namely, their low Eddington ratios $\lambda_{Edd}$ = L/ L$_{Edd}$ of 2-4$\times$ 10$^{-9}$ and their radiatively inefficient accretion flow (RIAF), it is likely that they both possess a rotating ionised disc in their nuclei, at a fraction of their Bondi radius. We thus know:
\begin{equation}
     \lambda_{Edd} = f(\dfrac{\dot{M}_B|_{M31} }{M_\bullet|_{M31}}) = f(\dfrac{\dot{M}_B|_{MW}}{M_\bullet|_{MW}})
\end{equation}
The accretion rates are proportional to their respective BH masses. This corresponds to a typical (2D) accretion rate at the Bondi radius:
\begin{equation}
     \dot{M}_B \propto 2\pi R_B \mu c_s, \mathrm{where\,} \mu  \mathrm{\,is \,the\, ionised\, gas\, surface\, density.}
\end{equation}
This means that we can assume the same plasma characteristics for the accretion around M31* and SgrA* (i.e. 
$\mu$, $c_s$, and plane thickness $h$), the accretion rate is 35 times larger for M31*:
\begin{equation}
    \dot{M}_B|_{M31}  = \left[ \dfrac{M_\bullet|_{M31}}{M_\bullet|_{MMW}} \right] \times \dot{M}_B|_{MW} = 35 \times \dot{M}_B|_{MW} 
\end{equation}
}
 {Last, the expected integrated signal can be written as\,:
\begin{equation}
    S \Delta V_{H30\alpha}|_{M31} = \dfrac{\epsilon_{H30\alpha}}{4\pi h D^2_{M31}} \mu^2 \pi R_{B}^2 \dfrac{c}{\nu_{obs}} ,
\end{equation}
where $\epsilon_{H30\alpha}$ the emissivity of ${\text{H30}\alpha}$ \citep{1995MNRAS.272...41S}, which varies weakly with the density. Assuming simple scaling relations we find\,:
\begin{eqnarray}
    S \Delta V_{H30\alpha}|_{M31} &=&  \left[\dfrac{D_{MW}}{D_{M31}}\right]^2 \left[ \dfrac{M_\bullet|_{M31}}{M_\bullet|_{MMW}}\right]^2 \times S \Delta V_{H30\alpha}|_{MW}\\
    &=& 0.13\times S \Delta V_{H30\alpha}|_{MW}
\end{eqnarray}}

{The expected ionised gas signal in M31 is 8 times lower than in the MW.
The 3$\sigma$ upper limit obtained here is 0.07 $\times$ the MW results. Therefore our upper limit is twice below what was expected. This means that the M31 nucleus has less ionised gas than SgrA* in proportion. }

\subsection{Continuum flux}
 Our upper limits are compatible with the synchrotron power law emission, derived from lower frequency measurements \citep{1992ApJ...390L...9C,1993ApJ...417L..61C,2017ApJ...845..140Y}. Given the absence of gas in the nucleus, no dust emission is expected.

\section{Discussion and conclusions}
\label{sect:disc}
The ALMA observations present an unprecedented sensitivity for cold molecular gas, further excluding its presence next to the black hole. Indeed, the modelling proposed by \citet{2007ApJ...668..236C} was predicting $10^4 M_\odot$ of CO(1-0), while our upper limit ($195 M_\odot$) is $50$ times smaller. While the principle of gas replenishment due to stellar wind from red giants in the excentric disc is reasonable, some details are probable incorrect. {Beside the small rotation pattern requirement that seems compatible with the modern estimates \citep[e.g.][]{2018ApJ...854..121L}, the values used for the stellar wind mass-loss rates  lie on the upper-side of the expected range and might be overestimated by a factor larger than $50$ (E. Josselin, priv. comm.). The modelling might still account for the formation of the blue nuclear cluster next to P3 \citep{2012ApJ...745..121L}, but the timescale to reconstitute a gas reservoir would then be longer than expected.}
{Alternatively, \citet{2007A&A...461..651D} also discussed that the central blue nucleus might be composed of  old stellar population of evolved blue horizontal-branch stars and of merger products, which would invalidate the need of gas inflow to form it.}

{Relying on simple scaling relations based on the \citet{2019Natur.570...83M} detection of the H30$\alpha$ recombination line next to SgrA*, we tentatively search this line next to M31*, with deep NOEMA observations. We excluded it at a 6$\sigma$ level. }
{In the optical, \citet{2013ApJ...762L..29M} have discovered an eccentric H$\alpha$ emitting disc in the M31 nucleus, of radius 0.7\,\arcsec. The presence of this optical recombination line traces the presence of some gas next to the nucleus. They estimate that this 2.7-pc-radius gas disc has a luminosity of $L_{H_\alpha}=(8.7\pm 1) \times 10^2\,L_\odot$. 
This luminosity corresponds to a recombination rate of $10^{48}$ photons per second.
One can discuss the possible mechanisms that might have ionised this gas.
Following \citet{1985ApJ...290..136J}, one typical planetary nebula emits $1.2 \, 10^{45}$ photons per second, while we do not expect more than a few planetary nebulae in this region \citep[e.g.][]{2013MNRAS.430.1219P}. The youngest stars present in this region belong to the A0-star cluster ($4200\,M_\odot$) next to the black hole in P3 \citep{2005ApJ...631..280B}. Their typical temperature $10\,000$\,K \citep{2000asqu.book..381D} excludes any significant amount of ionising photons. Assuming a black body emission, such an A0-star with a typical mass $2.5 \,M_\odot$ cannot produce more that $1.2\times 10^{38}$ recombinations per second, corresponding to a grand total of $2\times 10^{41}$ recombinations per second. We can thus exclude that stellar objects contribute to the ionisation of this central gas.}

{On 2006 January 6$^{th}$, \citet{2011ApJ...728L..10L} observed a murmur of M31* at a level of $4.3\times 10^{37}$ erg\,s$^{-1}$ in the frequency range $0.5-8$\,keV corresponding to a minimum of $2\times 10^{48}$ recombinations per second. The past relic activity of the central engine might well explain the excitation of this inner H$\alpha$ disc. Indeed, an episodic activity of the black hole is also supported by the work of \citet{Zhang-Foster2019}, based on X-ray gas,  who concluded that only an AGN episode half a million years ago could account for their observed line ratios. In addition, on probably different timescales,  \citet{2016MNRAS.459L..76P} also discussed signs of recent AGN activity found in possible Fermi bubbles, while \citet{2008MNRAS.388...56B} detected the possible relic of a past outflow.  Recurrent bursts from the central engine might account of the whole picture.
Hence, it is possible that the CO has been destroyed as proposed by \citet{2015ApJ...803...37B}, due to the black hole activity traced in X-ray. Moreover, one can quote the work of \citet{2007ApJ...661..203J} who show that the irradiation by an AGN can modify the atmosphere of red giant stars.

{Our upper limits and the detection of a weak H$\alpha$ disc by \citet{2013ApJ...762L..29M} support the evidence of a past activity of the AGN (e.g. kinetic jet feedback), which  might have significantly contributed to centrally quench this galaxy as discussed in \citet{2024MNRAS.527.5988H}.
}

}

\begin{acknowledgements}
This paper makes use of the following ALMA data: ADS/JAO.ALMA\#2019.1.00711.S. ALMA is a partnership of ESO (representing its member states), NSF (USA) and NINS (Japan), together with NRC (Canada), MOST and ASIAA (Taiwan), and KASI (Republic of Korea), in cooperation with the Republic of Chile. The Joint ALMA Observatory is operated by ESO, AUI/NRAO and NAOJ. \\
This work is based on observations carried out under project numbers W19BN and W01E  with the IRAM NOEMA Interferometer. IRAM is supported by INSU/CNRS (France), MPG (Germany) and IGN (Spain).”\\
This work benefited from the support of the {\em Action fédératrice ALMA-NOEMA} of Paris Observatory, and in particular the workshop organised within this project by Raphaël Moreno and Philippe Salomé. This project also got from support from the {\em Programme National Cosmologie et Galaxies}. Special thanks go to Eric Josselin and Franck Delahaye for useful information on stellar populations. Last, we are most grateful to the anonymous referee for the very constructive comments that
helped us to substantially improve the manuscript. 
\end{acknowledgements}

\bibliographystyle{aa}
\bibliography{biblio}

\end{document}